\documentclass{aastex}          

\usepackage{spr-astr-addons}    
\usepackage{url}

\begin{document}
%
\title{Relativistic positioning: four-dimensional numerical approach in Minkowski space-time.}

\shorttitle{<Relativistic positioning systems>}
\shortauthors{N. Puchades and D. S\'aez}

\author{Neus Puchades and Diego S\'aez}
\affil{Departamento de Astronom\'\i a y Astrof\'\i sica, 
Universidad de Valencia, 46100-Burjassot, Valencia, 
Spain.}


\begin{abstract}
We simulate the satellite constellations of 
two Global Navigation Satellite Systems: 
Galileo (EU) and GPS (USA). Satellite motions 
are described in the Schwarzschild space-time produced by 
an idealized spherically symmetric non rotating Earth. The trajectories are then 
circumferences centered at the same point as Earth.
Photon motions are described in Minkowski space-time, where there is a well known relation, 
\cite{col10}, between the emission and
inertial coordinates of any event. 
Here, this relation is implemented in a 
numerical code, which is tested and applied.
The first application is a detailed numerical four-dimensional analysis of the 
so-called emission coordinate region and co-region. In a second application,
a GPS (Galileo) satellite is considered as the receiver and
its emission coordinates are given by 
four Galileo (GPS) satellites. The bifurcation problem (double localization) in the 
positioning of the receiver satellite is then pointed out and discussed in detail.
\end{abstract}


\section{Introduction}
\label{intro}

Nowadays, in order to design an {\em operative} relativistic positioning system (RPS),
we need: (i) inertial coordinates ($x^{1},x^{2},x^{3},x^{4}$) labeling
events in an appropriate reference system, (ii) four satellites,
whose positions are known at any time, which
broadcast their proper times by means of electromagnetic signals,
(iii) detectors which receive the proper times from the four 
satellites at the same time. These times are the 
emission coordinates of the reception event, and
(iv) the transformation from emission to inertial coordinates,
which localizes the reception event in the inertial 
system of reference (relativistic positioning). In any RPS, 
some 4-tuples of emission coordinates may be received in two different positions
at two distinct coordinate times; namely, each of these 4-tuples
leads to two real and different reception events in Minkowski space-time. 
If one of these 4-tuples of proper times
is received by an observer, new information --apart from the emission 
coordinates-- is required to choose one of the two possible locations.
The study of this problem, in realistic cases, is one of the main goals of this paper

Although current positioning systems 
are based on Newtonian physics, relativistic post-Newtonian corrections are 
performed if necessary; however,
RPS should be based on relativistic principles from the beginning.
In the proposed schemes,
the proper times of four satellites (emission coordinates) are sent by means 
of electromagnetic 
signals to the receiver, whose inertial coordinates can be found
--from the emission ones-- by using fully
relativistic equations.  
Realistic four-dimensional (4D) implementations of the transformation derived by 
\cite{col10} require numerical calculations. See \cite{neu10} as a preliminary 
numerical application of this transformation. Here, numerical codes based on the 
same transformation have  
been designed and tested. Results obtained with them are described in next sections. 
The transformation 
from emission to inertial coordinates \citep{col10} in Minkowski space-time
--where it is asumed that the electromagnetic signals propagate--
uses the position of the four satellites 
when they emitted their proper times.
The circular motion of GPS and Galileo satellites has been simulated 
by using the Schwarzschild space-time created by an idealized 
spherically symmetric Earth (see next section); 
thus, the satellite positions may be 
calculated at any given time with a good enough accuracy.
Of course, the formalism must be extended to general relativity 
\citep{cad05,bin08,tey08,cad10,bun11,del11}
to include gravitational fields, accelerated frames, and so on,
in both satellite and  photon motions. 
Although a fully relativistic scheme should work without any reference to an Earth based 
coordinate system \citep{tar09}, realizations of this type of scheme
have not been implemented up to now.

In the study presented here, Earth and other possible obstacles to light propagation 
are not considered at all and, moreover, signals broadcast by the satellites 
may be detected at any distance by ideal receivers which have not a 
threshold for detection. In this way, only the space-time structure is
taken into account in our discussion. Finally, a spherically symmetric Earth is 
assumed and the satellites describe circular orbits whose centers 
must coincide with that of the Earth. These orbits are covered as it
must be done in Schwarzschild space-time.
Afterward, our study may be
generalized to include obstacles, realistic detectors and non symmetric 
distributions of mass inside Earth.

A coherent terminology for relativistic positioning may be
found in \cite{col10}, where concepts as emission coordinates, 
grid, emission coordinate region and co-region, and so on, 
were rigorously defined. Discussions in some previous papers 
\citep{col01,bah01,col02,rov02,bah03,col06a,col06b,col06c,col10b,col11}
led to synthesize the mentioned terminology and to define the foundations
of relativistic positioning.
A sketch of the emission region in 3D (sections of constant coordinate time) is presented  
in \cite{b6}. It is based on a system of  
three static satellites symmetrically placed in the vertexes 
of an equilateral triangle. These satellites broadcast their proper times
(emission coordinates) to receivers which live in a Minkowskian
space-time with one time dimension and two space ones.
Realistic 4D studies of emission coordinate regions and co-regions, 
with four dynamical satellites, have not
been performed yet. The first of these studies is developed here.

Quantities $G$, $M_{\oplus}$, $t$, and $\tau $ stand for the gravitation constant,
the Earth mass, the coordinate time, and the proper time, respectively.
Greek (Latin) indexes run from $0$ to $3$ ($1$ to $3$). Quantities 
$\eta_{\alpha \beta}$ are the covariant components of the Minkowski 
metric tensor. Units are defined in such a way that the speed of 
light is $c=1$.

The paper is organized as follows, in Sect.~\ref{sec-2}, the motion of the
GPS and Galileo Global Navigation Satellite Systems (GNSSs) is simulated,
the method used to obtain the emission (inertial) coordinates from the 
inertial (emission) ones in Minkowski space-time is described in 
Sect.~\ref{sec-3} (\ref{sec-4}). In the first (second) of these sections 
we use analytical (numerical) techniques. Section \ref{sec-5} 
contains numerical results. The emission coordinate regions (co-regions) defined in \cite{col10}
are described in Sect.~\ref{subsec-5.1} (\ref{subsec-5.2})
for various sets of
four satellites, and the possibility of finding
the position of a satellite by
using other four satellites --as emitters-- is discussed in Sect.~\ref{subsec-5.3}.
Finally, Sect.~\ref{sec-6} contains a general discussion and 
some comments about perspectives. 

\section{The satellites}
\label{sec-2}

Two GNSSs are considered: GPS in USA and Galileo (under construction) in the EU. 
They provide the spatial coordinates and the universal time of any event on Earth. 
Position coordinates are calculated thanks to information received from satellites 
into orbit around Earth. The GPS constellation has $n_{s}=24$ satellites 
and it is arranged in six 
different orbital planes (four satellites per plane), each of them inclined 
an angle $\alpha_{in}=55$ degrees 
with respect to the equator. 
To obtain around two orbits per day, the satellites are placed at an altitude $h=20200 \ Km$. 
The Galileo constellation is composed by 27 satellites 
($n_{s} = 27$), located in three equally spaced orbital planes 
(9 equally spaced satellites 
in each plane). The inclination of these planes is $\alpha_{in}=56$ degrees 
and the altitude of the circular 
orbits is $h=23222 \ Km$; thus, the orbital period is about 14h.

Satellites move in Schwarzschild space-time to take into account Earth gravity. 
The trajectory of any satellite is assumed to be a circumference of radius $R$,
which has the same center as Earth. In Schwarzschild space-time, 
the angular velocity on these circumferences is
$\Omega = (GM_{\oplus}/R^{3})^{1/2} $, so
in a coordinate system attached to 
Earth center, the coordinates of a given 
satellite $A$ may be written as follows:
\begin{eqnarray}
x^{1}_{A} &=& R \, [\cos \alpha_{A}(\tau) \cos \psi + \sin \alpha_{A}(\tau) 
\sin \psi \cos \theta] \nonumber \\ 
x^{2}_{A} &=& -R \, [\cos \alpha_{A}(\tau) \sin \psi - 
\sin \alpha_{A}(\tau) \cos \psi \cos \theta] \nonumber \\
x^{3}_{A} &=& -R \sin \alpha_{A}(\tau) \sin \theta \nonumber \\
x^{4}_{A} &=&  \gamma \tau   \ .
\label{satmot1}
\end{eqnarray} 
The factor $\gamma $ calculated up to first order in $GM_{\oplus}/R$ is given by the relation 
\begin{equation}
\gamma = \frac {dt}{d\tau} = \Big( 1 - \frac {3GM_{\oplus}}{R} \Big)^{-1/2} \ ,
\label{ttau} 
\end{equation}  
and the angle
\begin{equation}
\alpha_{A}(\tau) = \alpha_{A0} - \Omega \gamma \tau
\label{satmot2} 
\end{equation}  
localizes the satellite on its trajectory. Finally, $\theta $ and $\psi $ are Euler 
angles associated to two systems of spatial axis: the axis ($x^{1},x^{2},x^{3}$) 
trivially associated to the standard angular Schwarzschild coordinates, 
and a second set of axis ($x^{\prime 1},x^{\prime 2},x^{\prime 3}$), 
which is chosen in such a way that  
($x^{\prime 1},x^{\prime 2}$) coincides with the orbital plane 
containing the trajectory of the satellite under consideration. Angle $\theta = 2\pi - \alpha_{in}$
is the same for all the satellites of a given GNSS,
whereas angle $\psi $ takes on the values $\psi = (2\pi / n_{so})(j-1)$, where 
$n_{so} $ is the number of satellites per orbital plane and the 
natural number $j$ labels these planes. 
Evidently, angle $\psi$
is the same for all the satellites of a given orbital plane. 
For any satellite, angles $\theta $, $\psi $ and $\alpha_{A0}$ are constant.
The last angle defines the position of satellite $A$ at $\tau = x^{4} = 0$.
This angle may be arbitrarily chosen for a satellite in each orbital plane and, then, 
the remaining $\alpha_{A0}$ angles may be fixed in 
such a way that all the satellites are equally spaced on their 
common trajectory.

\section{From emission to inertial coordinates}
\label{sec-3}

Events of interest are always simultaneous observations of 
four satellites. The inertial coordinates of one of these events 
are denoted $x \equiv (x^{1},x^{2},x^{3},x^{4}) \equiv (\vec{x},t)$. The emission 
coordinates are the four proper times, $\tau^{A}$, 
codified in the satellite 
signals, where index $A$ numerates the satellites.  

The coordinates ($x^{1},x^{2},x^{3}$) of the satellite $A$, at emission time 
$\tau^{A}$, are denoted $\gamma_{A}$. Since the 
world lines of the satellites are known, quantities $\gamma_{A}$ 
may be calculated for arbitrary proper times.  
Vectors $e_{a} = \gamma_{a} - \gamma_{4} $ (with index $a$ running
from 1 to 3) define the relative 
position between satellite $a$ and satellite $4$, which is 
hereafter the emitter of reference. The numeration of the satellites 
and, consequently, the choice of the fourth satellite 
are arbitrary. We may say that vectors $e_{a} $ 
define the internal satellite configuration 
at emission times. There are inertial coordinates associated to
times $\tau^{A}$, if and only if, the so-called emission-reception conditions, 
\cite{col10}, are satisfied. These conditions may be written as follows: 
\begin{equation}
\eta_{\alpha \beta} e_{a}^{\alpha} e_{a}^{\beta} > 0, \,\,\,\,     
\eta_{\alpha \beta} (e_{a}^{\alpha}-e_{b}^{\alpha} ) 
(e_{a}^{\beta} - e_{b}^{\beta} ) > 0  \ , 
\label{e_r_c}
\end{equation}  
for any value of indexes $a$ and $b$ which run from 1 to 3. 
If these conditions are satisfied, 
we may look for the inertial coordinates.

In Minkowski space-time, 
the general transformation from emission to inertial 
coordinates was derived in \cite{col10}; it may be written as follows:
\begin{equation}
x = \gamma_{4} + y_{\ast} - \frac {y_{\ast}^{2} \chi} 
{(y_{\ast} \cdot \chi)+\hat{\epsilon}\sqrt{(y_{\ast} \cdot \chi)^{2}-
y_{\ast}^{2} \chi^{2}}}   \ ,
\label{emisin}
\end{equation}  
where vectors $\chi $ and $y_{\ast} $ may be calculated from
$e_{1} $, $e_{2} $, and $e_{3} $ (internal satellite 
configuration). The configuration vector $\chi = \ast(e_{1}\wedge 
e_{2}\wedge e_{3}) $ (dual of a double exterior product)
is orthogonal to the hyperplane containing the four $\gamma_{A}$
emission events. Vector $y_{\ast} = (\xi,H)/(\xi \cdot \chi) $,
where $(\xi,H)$ stands for the interior product,
may be calculated from any arbitrary vector $\xi $ satisfying the 
condition $\xi \cdot \chi \neq 0$ and from the bivector
$H = [(e_{A} \cdot e_{A})/2] E^{A} $, where 
$ E^{1} = \ast(e_{2} \wedge e_{3}) $,
$ E^{2} = \ast(e_{3} \wedge e_{1}) $, and  
$ E^{3} = \ast(e_{1} \wedge e_{2}) $.

The above transformation is the solution of a 
system of four equations (hereafter the main system). Each equation expresses that 
the distance from $\gamma_{A}$ to $x$ vanishes; so two  
types of solutions appear. The first type corresponds to signals 
{\em emitted from the satellites} at times $\tau^{A}$ and simultaneously received 
at position $\vec{x}$ and time $t$ (emission or past-like solutions). The second type describes 
a signal emitted from position $\vec{x}$ and time $t$ and {\em received 
by the satellites} at times $\tau^{A}$ (reception or future-like solutions). 
In any RPS we are only 
interested in the first type. Hereafter, solutions of this type are also 
called {\em positioning solutions}.

For $\chi^{2} \neq 0$, there are two sets 
of inertial coordinates corresponding to $\hat{\epsilon}=+1$ and 
$\hat{\epsilon}=-1$. Moreover, for $\chi^{2} < 0$, 
only one of the two sets of inertial coordinates corresponds to a 
positioning solution. In the case $\chi^{2} > 0$, the number of positioning 
solutions may be either two or zero, in the first case, 
there are two different receptors 
(located at different places),
which would receive the same four emission times from the same 
satellites. In the second case, there are two future-like solutions (zero
positioning solutions). Finally, for $\chi^{2} = 0$ there is only a positioning 
solution corresponding to $\hat{\epsilon}=+1$.

In case $\chi^{2} < 0$, the positioning solution satisfies the condition 
$t_{A} - t < 0 $ for any $A$, whereas the inequalities $t_{A} - t > 0 $
are satisfied for the future-like solution. Since the satellite world lines are 
known, the inertial coordinate $t_{A} $ may be calculated at any
proper time $\tau^{A} $ and, consequently, the sign of $t_{A} - t $ may          
be used to identify the positioning solution.

If one has a bifurcation problem with $\chi^{2} > 0$ and two positioning solutions $x_{1}$ and $x_{2}$
\citep{sch72,abe91,cha94,gra96},
the receiver should have a criterion to select its true inertial 
coordinates. One of these criteria --proposed in \cite{b5,b6}-- is as follows: 
consider a conical surface with the receiver at the vertex which contains three of the four 
satellites and then, take 
either one sign of 
$\hat{\epsilon} $ or the opposite one
depending upon whether the fourth satellite 
is inside or outside the cone, respectively: of course,
the receiver should have devices to measure angles.
Since the $x^{4} $ coordinate times of the two
positioning solutions are different, the receiver may select the 
true positioning solution by using a clock. Only 
the coordinate time of the true solution will be identical 
(close enough taking into account the clock accuracy and possible positioning errors) to 
that given by the receiver clock.

\section{From inertial to emission coordinates}
\label{sec-4}

Given the inertial coordinates, $x^{\alpha}$, of an event in Minkowski space-time,
its emission coordinates, $\tau^{A}$, may be numerically calculated. 
Let us now describe the method we have implemented to perform this calculation.
Since emission and reception events must be on a null geodesic,
we can write the following algebraic equations  
\begin{equation}
\eta_{\alpha \beta} [x^{\alpha} - x^{\alpha}_{A}(\tau^{A}) ]  
[ x^{\beta} - x^{\beta}_{A}(\tau^{A}) ] = 0   \ , 
\label{inemis}
\end{equation}    
in which the proper times $\tau^{A}$ are the unknowns. The
solution of these equations are the emission coordinates $\tau^{A}$.
This solution may be easily obtained                    
by using the Newton-Raphson method \citep{pre99} plus Eqs.~(\ref{satmot1})
and (\ref{satmot2}). 

   After obtaining the four emission coordinates $\tau^{A}$, we can use 
Eq.~(\ref{emisin}) to recover the inertial coordinates we had
initially chosen. We use multiple precision in the code designed to
solve Eq.~(\ref{inemis}) and also in the numerical calculations based on 
Eq.~(\ref{emisin}). If a precision of forty digits is required, 
we have verified that the parameter fixing the precision of the 
Newton-Raphson code may be adjusted to 
recover 39 digits after computing the initial inertial coordinates 
with Eq.~(\ref{emisin}). This test ensures that  
our numerical calculations leads to very 
accurate emission (inertial) coordinates starting from 
inertial (emission) ones. In other words, we have very accurate codes 
to calculate inertial coordinates from (\ref{emisin}), as well as to 
get emission coordinates by solving Eq.~(\ref{inemis}). The 
second of these calculations --based on Newton-Raphson method-- 
is more time consuming.

\section{Numerical results}
\label{sec-5}

In this section, the emission region and the co-region --defined in
\cite{col10}-- 
are numerically studied for the first time, in the case of realistic satellite
configurations (GPS and Galileo).

Numerical methods and codes based on the transformations between
inertial and emission coordinates have been described in 
Secs.~\ref{sec-3} and~\ref{sec-4}.

The space $\mathcal{R}^{4} $ containing all the 4-tuples of proper times 
is called {\em grid}.
Given an arbitrary point of the grid 
($\tau^{1}$,$\tau^{2}$,$\tau^{3}$,$\tau^{4}$), the question is:
would these proper times be received in some point of Minkowski space-time?
In other words, are there positioning solutions of the main system for these proper times?
If affirmative, the points in Minkowski space-time (receivers) would belong to the 
so-called emission region and the chosen grid point to the emission co-region, 

The emission region and its co-region are 4D spaces and, consequently,
graphic descriptions require the study of a set of appropriate 3D sections.
In the emission region (co-region) the chosen 3D sections are characterized by the 
condition $x^{4} = constant$ ($\tau^{4} = constant$).

A few words about the figures are worthwhile to 
ensure a right intuitive interpretation.

In this paper, the
HEALPIx ({\it hierarchical equal area isolatitude pixelisation
of the sphere}) package \citep{gor99} is used to 
depict the figures. This pixelisation was designed to 
construct and analyze maps of the cosmic microwave background.
It is useful to display any quantity depending on the
observation direction (pixel).
The number of pixels is $12 \times N_{side}^{2}$, where 
the free parameter
$N_{side} $ takes on even natural values. 
In Figs.~\ref{figu1} and~\ref{figu2} corresponding to the 
emission region (\ref{subsec-5.1}), we will take $N_{side} =16$ ($3072$ pixels),
whereas in the Figs.~\ref{figu3} to~\ref{figu5} of the co-region
(\ref{subsec-5.2}), we will use
$N_{side} =32$ ($12288$ pixels).
The angular area of any pixel is $\sim 13.43 $ ($\sim 3.36 $)
squared degrees for $N_{side} =16$ ($N_{side} =32$).
In the case of $N_{side} =16$ ($N_{side} =32$) the pixel is close 
to sixty four (sixteen) times the mean angular area of the full moon,
but its shape is not always the same (as it is appreciate in 
the figures). Pixels are more elongated in the polar zones.

Finally, the pixelised sphere is shown by using the mollwide 
projection, in which, the frontal hemisphere is represented in the 
central part of the figure. The opposite semi-sphere is divided
in two zones which are displayed in the lateral parts
of the same figure, whose edges represent the same 
back semi-meridian.

\subsection{Emission region structure}
\label{subsec-5.1}

The emission region of four satellites is  
the zone of the space-time where proper times from them 
(emission coordinates) may be received.
In Minkowski space-time, any point, whose emission 
coordinates (see section \ref{sec-4}) satisfy the relation
$\tau^{A} \geq \tau^{A}_{in} $ for the four satellites, 
belongs to the emission region; evidently, 
$\tau^{A}_{in} $ is the time at which the satellite $A$ started to
emit. Signals from the four satellites will reach any
position ($x^{1},x^{2},x^{3}$) --whatever its distance to Earth may be--
for a certain value of the coordinate time. The larger the 
distance to Earth, the greater this time.

Two types of reception events are distributed in the emission region.
The first type is characterized by the condition 
$\chi^{2} \leq 0$ and, consequently, there is only a positioning solution 
(hereafter, single positioning); however, the second type
corresponds to $\chi^{2} > 0$ and, in such a case, there are pairs 
of events corresponding to the same emission coordinates
(hereafter, double positioning or bifurcated location).
An additional criterion is necessary for positioning (see section \ref{sec-3}). 
We are interested in the 
distribution of these types of events inside the emission region.

In order to perform an exhaustive study of various emission 
regions we proceed as follows:
(1) a reception event is selected. It occurs
at coordinate time $t_{R} $ in a point on 
the Earth surface with coordinates $(x_{e}^{1},x_{e}^{2},x_{e}^{3})$. From the 
coordinates ($x_{e}^{1},x_{e}^{2},x_{e}^{3},t_{R}$), the emission coordinates 
and the quantity $\chi^{2} $ may be calculated (see Secs.~\ref{sec-3} and ~\ref{sec-4}). 
Since the 
relation $\chi^{2} < 0$ is always satisfied on Earth, 
the selected event corresponds to a single 
positioning. In all the cases considered along the paper, coordinates
$(x_{e}^{1},x_{e}^{2},x_{e}^{3})$ are always the same, and they correspond to
a point on Earth with the spherical coordinates $\theta_{e} = 60^{\circ} $ and
$\phi_{e} = 30^{\circ} $. Three 4-tuples of satellites (labeled 1, 2, and 3) are studied.  
Finally, various times $t_{R} $ covering a period of the Galileo satellites 
are considered for 4-tuple 1; (2)  
the hypersurface $x^{4} = t_{R}$ (emission region section) is studied. 
In order to do that, the HEALPIx pixelisation 
is used to define
a set of $3072$ directions (see Sect.~\ref{sec-5}). A straight line starting from
the selected point is associated to each direction; (3) $N_{d} $ equally 
spaced points are defined on each straight line. These points cover a 
certain maximum distance, $L_{max} $, measured from the selected 
central event. Each of these points --plus the fixed time $t_{R} $-- 
is a possible reception event whose
emission coordinates may be found by using
the Newton-Raphson algorithm (see section \ref{sec-4}) and, (4) from the 
emission coordinates (which allow us to find the satellite positions) 
we may determine the sign of $\chi^{2}$ to know 
whether we are concerned either with a single ($\chi^{2} \leq 0$) or with a double 
($\chi^{2} > 0$) positioning solution. 

For a given HEALPIx direction, the following cases have appeared:
(i) $\chi^{2} < 0$ from the chosen point on Earth ($L=0$) to a 
certain distance $L_{-} $ (where $\chi^{2}$ vanishes),
and $\chi^{2} > 0$ from $L_{-} $ to $L_{max} $, and
(ii) $\chi^{2} < 0$ from $L=0$ to $L_{max} $, which means that 
either $\chi^{2} $ does not vanish along this direction or $L_{-} > L_{max} $. 
We may then represent length $L_{-} $
on a pixelised sphere by using both the mollview projection (see Sect.~\ref{sec-5})
and an appropriate color bar; thus, the pixel color 
gives the distance $L_{-} $ for the corresponding direction. 
In this way, 
whatever the chosen event on Earth may be, we are displaying a 
surface which separates single from double positioning solutions 
for the fixed hypersurface $x^{4} = t_{R}$. Single solutions 
are located either on the mentioned surface or at the 
same side --with respect to the surface-- as the Earth point 
chosen as a center. Double positioning solutions are all located 
at the other side of the surface, which is hereafter called the 
separating surface. 

In some papers \citep{col10,b6}, the parts of the emission coordinate region 
characterized by the conditions $\chi^{2} < 0$, $\chi^{2} = 0$, and  $\chi^{2} > 0$,
are denoted $C_{s} $, $C_{l} $, and $C_{t}$, respectively. 
Hence, our separating surface, the region containing the Earth point
playing the role of a center, and the complementary region with 
double solutions are the intersections of the hypersurface $x^{4} = t_{R}$
with $C_{l} $, $C_{s} $, and $C_{t}$, respectively.

Figure \ref{figu1} shows the separating surfaces
corresponding to three 4-tuples of satellites belonging to the
Galileo constellation. The value of $t_{R}$ has been 
arbitrarily chosen for each 4-tuple. 
Figure \ref{figu2} displays the separating surfaces for
the 4-tuple 1 considered in the top panel of Fig. \ref{figu1}.  
Six new values of $t_{R}$ (six hypersurfaces) have been considered in 
this Figure (one in each panel). They cover 
--together with the $t_{R} $ value of the top panel of Fig. \ref{figu1}--
a period of the Galileo constellation.      
In both Figures, 
the maximum distance has been chosen to be $L_{max} = 10^{5} \ Km $,
the white pixels contain the directions pointing 
towards the four satellites 
at the chosen emission times, the garnet region corresponds to directions 
with $L_{-} > L_{max} $, and the remaining pixels ($L_{-} < L_{max} $) 
are colored according to the color bar appearing in each panel (numbers in the bar are
values of $L_{-} $ given in Kilometers).
It is easily observed that: (i) 
all the satellites seem to be included in a blue region of influence, (ii)
various satellites may be located in the same region, and (iii)
inside the blue zones, the satellites may be located in the central part
as well as  
in the zones close to garnet regions. Of course, the 
distribution and positions of the blue regions around the 
satellites depend on the relative positions of the four satellites
among them and with respect to the receiver (at emission times). 
In the directions of the garnet regions,
our study has been repeated for a larger distance 
$L_{max} = 3 \times 10^{5} \ Km $, 
thus it has been verified that only 
for a very small number of pixels (located close to the non garnet 
region), a $L_{-} $ value satisfying the inequalities 
$10^{5} < L_{-} < 3 \times 10^{5}$ has been found. Hence,
in new Figures corresponding to $L_{max} = 3 \times 10^{5} \ Km $, these pixels 
would not be anymore in the garnet region, but in the 
complementary one. Note that the new $L_{max} $ is close to
the distance from Earth to moon and, consequently, 
if the moon is (is not) located in the garnet region of the chosen 
satellites, positioning on 
its surface would be single (double).
In the case of double positioning,  
additional measurement would be necessary to choose one of the two possible
localizations. Positioning at these large distances --from the satellites-- is
theoretically possible, although technical problems 
would arise (weak signals, large positioning errors 
due to uncertainties in the satellite trajectories, and so on).

Various sections, $x^{4} = constant$,
of some emission coordinate regions  
are represented in Figs.~\ref{figu1} and~\ref{figu2}. 
Let us now consider the intersections of the 
celestial spheres of these Figures with planes
containing meridians; thus, 2D sections of the 
emission coordinate regions are found. 
It may be easily verified that, the structure of the resulting 2D sections 
is analogous to that of the sketch displayed in \cite{b6}, 
but as expected, our realistic sections 
are less symmetric. In both cases we may distinguish the 2D sections of the 
so-called central 
region with single positioning ($C_{s} \cup C_{l}$), and complementary 2D 
sections of $C_{t} $, which contain all the double locations.

From Figs.~\ref{figu1} and~\ref{figu2}, it follows 
that close enough to Earth, up to distance of the order of $10^{4} \ Km $
from the surface, positioning 
is single (the exact value of $L_{-}$ depends on direction).
However, for larger distances, 
the positioning may be either single or double and, consequently,
if a Galileo (GPS) satellite is positioned by using four GPS (Galileo)
satellites, intervals of single and double positioning are expected
(see below).

\subsection{Co-region structure} 
\label{subsec-5.2}

In order to study the co-region, 
four steps are followed: (1) a point in the grid with
coordinates
($\tau_{e}^{1},\tau_{e}^{2},\tau_{e}^{3},\tau_{R}$) 
--which are the emission coordinates of the event selected 
in the last subsection ($\chi^{2} < 0$ $\Longleftrightarrow$ single positioning)-- 
is chosen to play the role of central point,
(2) time $\tau^{4} = \tau_{R}$ is fixed (section of the 4D co-region), 
(3) times 
$\tau^{1}$, $\tau^{2}$, and $\tau^{3}$ are varied along 
straight lines starting from the central point and following the 12288 directions 
corresponding to a HEALPIx pixelisation (see Sect.~\ref{sec-5}), and (4) each line is uniformly covered 
by a set of $N_{d}$ points, and quantity 
$\chi^{2}$ is calculated at each point to classify the associate 
reception event (single or double positioning).    
                                                
Points on the straight lines are characterized by the parameter 
$\lambda= [(\tau^{1}-\tau_{e}^{1})^{2} + (\tau^{2}-\tau_{e}^{2})^{2} +
(\tau^{3}-\tau_{e}^{3})^{2}]^{1/2}$. The values of this 
parameter are given in seconds. Along each straight line, 
it may be numerically verified that 
the emission-reception conditions (\ref{e_r_c})
are only satisfied from $\lambda =0$ to a certain value 
$\lambda_{max} $. Once quantity $\lambda_{max} $ has been 
numerically determined for each HEALPIx direction,
the number of points $N_{d}$ and the separation between two
neighboring ones may be appropriately chosen to cover the segment 
limited by points $\lambda =0$ and $\lambda_{max} $ for
any direction.

We have verified that, along any HEALPIx direction, quantity  
$\chi^{2}$ is negative from $\lambda = 0$ to a                                       
certain $\lambda_{-} $ (where it vanishes), and positive 
from $\lambda_{-} $ to $\lambda_{max} $. Nevertheless, in 
the interval ($\lambda_{-}$,$\lambda_{max} $), where the condition  
$\chi^{2} > 0 $ is satisfied, two cases may be distinguished:
(i) there are two positioning solutions, and (ii) there are two
future-like solutions (no positioning ones). 

In order to display the properties of the co-region, two panels 
are presented for each $\tau^{4} = \tau_{R}$ hypersurface.
An appropriate HEALPIx pixelisation, the mollwide 
projection, and color bars are used in both panels. 
In one of these panels the quantity 
represented in each pixel is $\lambda_{-} $. Hence, we are 
representing a surface which surrounds the zone where 
the positioning solution is unique. This zone is hereafter referred as to the  
single valued co-region.
In the other panel, the color bar is used to show the difference 
$\lambda_{max} - \lambda_{-} $ for 
the pixels corresponding to directions with two positioning solutions,
whereas the gray zone displays the pixels where no positioning solutions
exist for $\lambda > \lambda_{-} $. In other words,
the non gray part gives, for each pixel, the width of the zone where there are 
double positioning solutions, which is located outside the 
single valued co-region. This second external zone is hereafter called
the double valued co-region. 
Fig.~\ref{figu3} shows the single (left panels) and double 
(right panels) valued co-regions for the same satellite 4-tuples as in 
Fig.~\ref{figu1} and for arbitrary values of $t_{R} $. 
In Fig~\ref{figu4} (\ref{figu5}) we 
display single (double) valued co-regions for
the 4-tuple 1 and various times. The time $t_{R} $ of the top panel
of Fig.~\ref{figu3} and the six times appearing in the panels 
of Figs~\ref{figu4} and ~\ref{figu5} (the same considered 
in Fig.~\ref{figu2}), cover a period of the 
Galileo satellites.
The central point in the co-region and the 
equation of the hypersurface $\tau^{4} = \tau_{R}$ are obtained 
from the inertial coordinates of the chosen event on Earth by 
using the Newton-Rhaphson method. Since coordinates $(x_{e}^{1},x_{e}^{2},x_{e}^{3})$
are the same in all cases (see above), only coordinate $t_{R} $ is given  
on top of each panel.
In all cases, we find an internal single valued region 
partially surrounded by a external double valued one. 

In the study of the emission region, the Newton-Raphson method must be 
applied for each direction of the HEALPIx pixelisation, whereas in the 
co-region, this method is only used at emission time. This makes
the study of the region more time consuming. By this reason, the chosen HEALPIx
realizations of the region have less directions than those of the
co-region (see Secs.~\ref{sec-5},~\ref{subsec-5.1} and~\ref{subsec-5.2}).
Anyway, all the maps have a good enough angular resolution. 

\subsection{Positioning GPS (Galileo) satellites with the Galileo (GPS) constellation} 
\label{subsec-5.3}

It has been argued \citep{tar09,col06b,col06c,col10b} that, in order to define an intrinsic coordinate 
system by using the proper satellites (with no reference to Earth), and also to 
measure the gravitational field in 
the region where the satellites move (gravimetry), it is necessary the interchange 
of information among these satellites and with the receiver. Let us consider 
the most simple information exchange, in which 
one GPS (Galileo) satellite is considered as a receiver to be positioned with
the emission coordinates broadcast by four satellites of the Galileo (GPS) constellation.
On account of our study of the emission region presented in section \ref{subsec-5.1}
(see the last paragraph),
it is evident that, in most cases, double positioning is expected to appear 
on the world line of the 
satellite playing the receiver role.

We have first considered two cases in which a Galileo satellite (receiver)
is positioned by using emission coordinates from four GPS satellites (emitters).
In each case, the world line of the receiver is known. On account of this fact, 
the following steps allow us to get the single and double 
receiver positions: (1) given the 
inertial coordinates of a point on the receiver world line, the Newton-Raphson method
(see section \ref{sec-4}) is used to get the emission coordinates, (2) 
from the resulting emission coordinates, quantity 
$\chi^{2}$ is computed to determine the positioning character
(single or double location) and, (3)
if the positioning is double, we use Eqs.~(\ref{emisin})
to get both positions, one of them is always on the circular orbit of the 
receiver satellite, and the other one is outside this trajectory. 
Results are presented in Figs.~\ref{figu6} and~\ref{figu7}. In order
to build up these Figures, a special
procedure has been designed which allows us to display the motion of both 
the receiver satellite
along its circular orbit and the associated point (if it exists) on its trajectory. 
These Figures are 3D representations of the trajectories in the 
$(x,y,z) \equiv (x^{1},x^{2},x^{3})$ space. 
$N_{c} = 7200$ equally spaced points are considered on the 
satellite circumference.    
Colors are used 
to follow the motion of the positioning solutions along their
trajectories. In Figs.~\ref{figu8} and~\ref{figu9}, the same techniques are 
used, and the same representations are displayed, but these Figures
correspond to a GPS receiver satellite
positioned from the emission coordinates broadcast by four Galileo satellites.

In Figs.~\ref{figu6} to~\ref{figu9}, 
the motion along any trajectory is dextrogyre.
Single positioning solutions are represented by red points; hence,
these points are always on the circular orbit of the receiver satellite.
An initial point is arbitrarily chosen.
Four subgroups of $N_{c} / 4$ points have been selected. The double 
positioning solutions are displayed by using the black, fuchsia, 
dark blue, and light blue, in the first, second, third and fourth
subgroups, respectively. The first point of the first 
subgroup is the chosen initial point, and points and subgroups 
are ordered in the sense of growing times (dextrogyre sense in Figures).
According to these criteria, the initial point 
is marked by a red star 
on the circular trajectory for single positioning 
(as in Fig.~\ref{figu6} and~\ref{figu8}), 
and by two black stars for double positioning 
(as in Fig.~\ref{figu7} and~\ref{figu9}).
The final point coincides with the initial one on the circumference, and
the associated point (if it exists) will be evidently represented by a 
light blue star (as in Figs.~\ref{figu6},~\ref{figu7}, and~\ref{figu9}). 
Since the periods of the GPS and Galileo satellites
are different, the character of the positioning (single or double valued)
at the initial and final points may be different, it is due to the 
fact that, though these points coincide on the circumference, the 
locations of the four emitter satellites is different at the initial
and final situation and, consequently, the positioning character 
may be distinct. 
Moreover, as points on the circumference --located inside a double valued 
region of any color-- tend to possible points separating this region of
contiguous single valued ones, the associated point lying outside the circumference 
tends to infinity and, consequently, the two positioning solutions 
tend to a unique real one.
Changes of color at the initial point appear as a result of 
differences between the positioning character at the initial
and final situations (see above).
In practice, the initial point never separates single and double valued zones and,
consequently, if there is a final positioning point outside the circumference,
it does not tends to infinite as the corresponding point on the 
circumference approaches the initial one.
Taking into account all these criteria and comments,
Figs.~\ref{figu6} to~\ref{figu9} may be easily understood. The brief 
but illuminating description given in the Figure captions
deserves attention.
 
\section{Discussion and prospects} 
\label{sec-6}

This paper has been essentially devoted to the study of the 
bifurcation problem (double positioning) in relativistic 
positioning systems. In order to develop this study, 
the following approach has been used:
photons move in a 4D Minkowskian space-time, and
satellites evolve in a 4D Schwarzschild space-time 
associated to a spherically symmetric Earth. This procedure 
takes into account the effects of the Earth gravitational field
on the clocks; for example, we have verified the well 
known fact that, for the GPS configuration, satellite clocks run more rapid than 
clocks at rest on Earth by about 38.4 microseconds per day. Our simulation  
of the GPS and Galileo constellations is accurate enough
for many estimations and, in particular, to study bifurcation.
Double positioning situations are identified on the emission coordinate
region and co-region, as well as on the orbits of some 
satellites, which are considered as receivers to be localized
with the help of four emitters. Methods to choose the true 
location in the case of bifurcation are proposed and 
discussed. Some open problems are pointed out (see below).

\begin{figure*}[tb]
\begin{center}
\resizebox{0.7\textwidth}{!}{%
\includegraphics{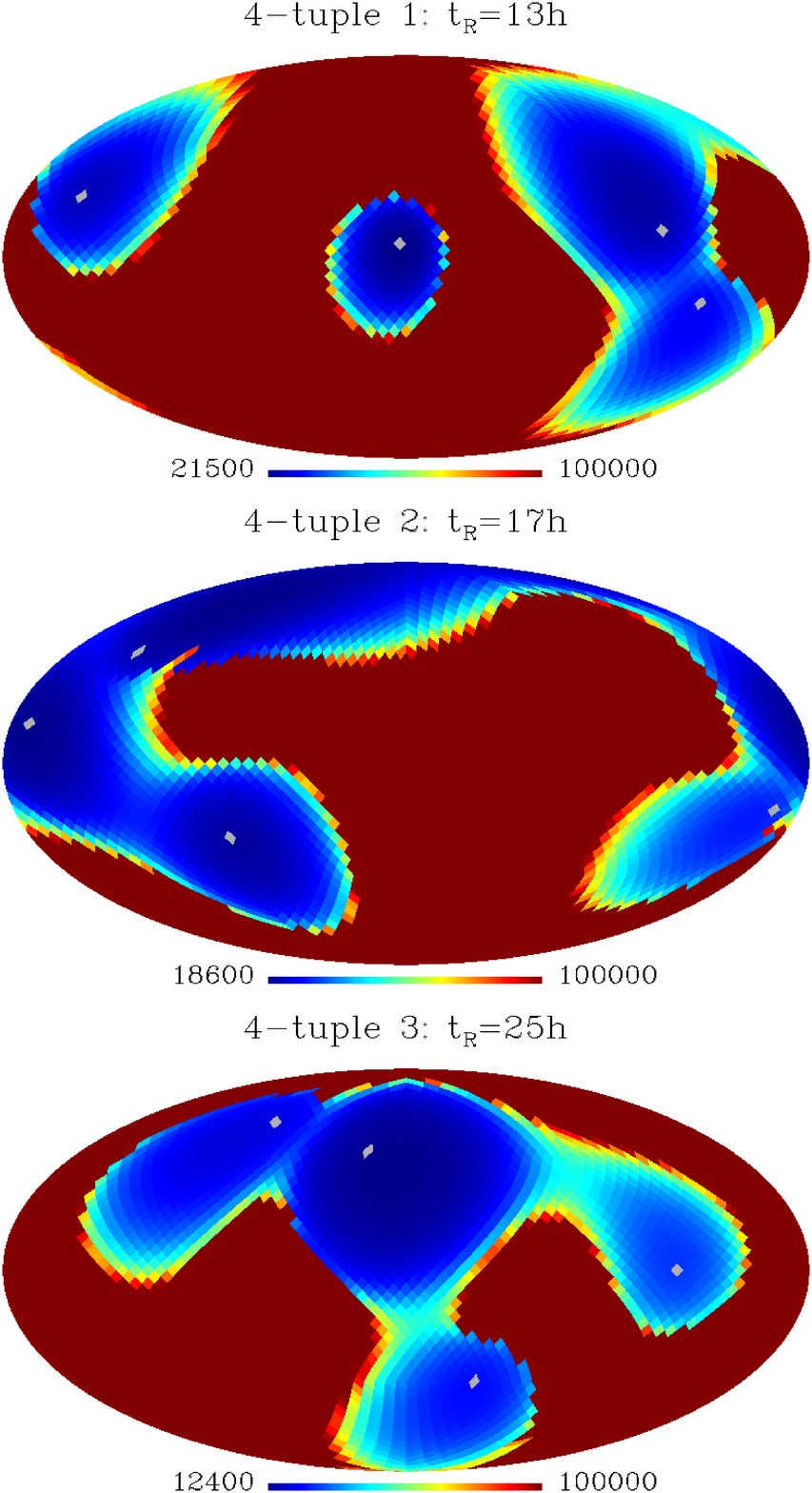}
}
\caption{All panels are HEALPIx mollwide representations of the distance 
$L_{-}$ in Kilometers. At this distance from the center, quantity $\chi^{2}$ vanishes
for each direction in the hypersurface 
$x^{4} = t_{R}$. This surface separates single from double valued regions.
The times $t_{R} $ and the 4-tuple of satellites used for positioning are
given above each panel.}
\label{figu1} 
\end{center}      
\end{figure*}

\begin{figure*}[tb]
\begin{center}
\resizebox{0.9\textwidth}{!}{%
\includegraphics{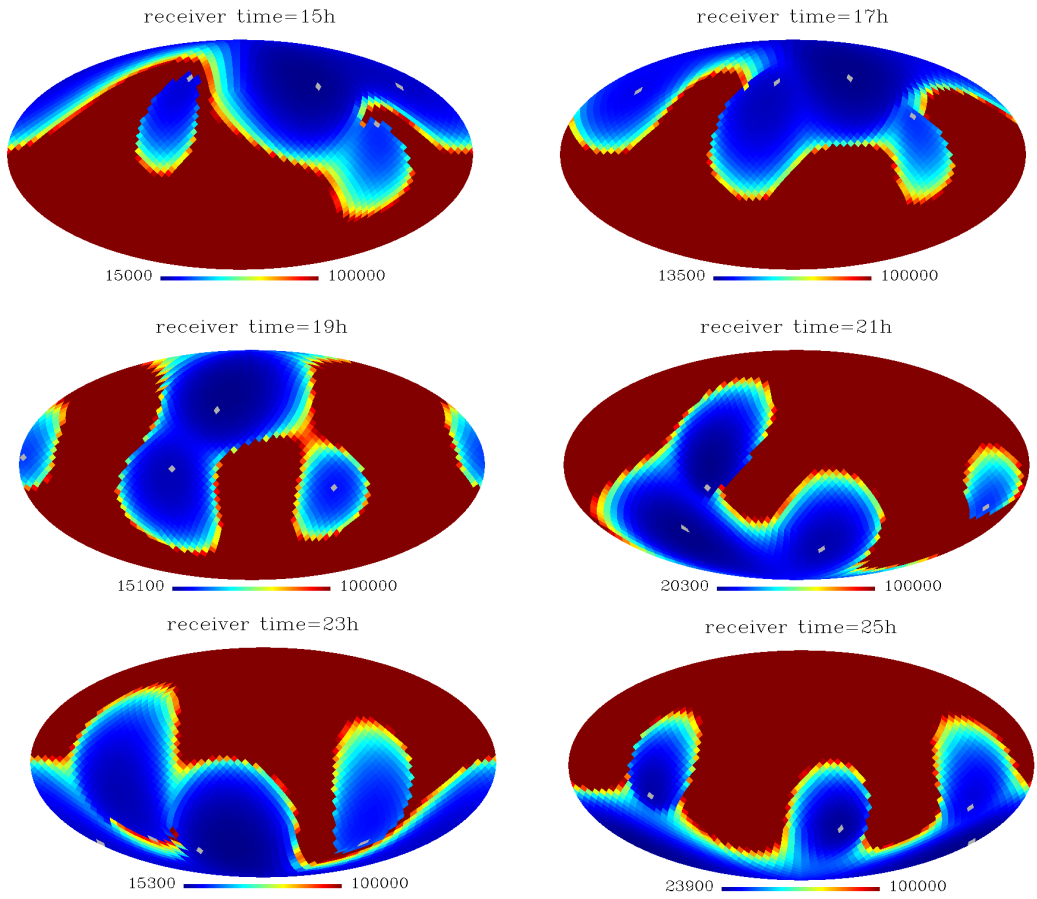}
}
\caption{Same as in Fig.~\ref{figu1} for the 4-tuple 1 
(top panel of that Figure) and for the $t_{R} $ times displayed 
above each panel.}
\label{figu2} 
\end{center}      
\end{figure*}

\begin{figure*}[tb]
\begin{center}
\resizebox{0.9\textwidth}{!}{%
\includegraphics{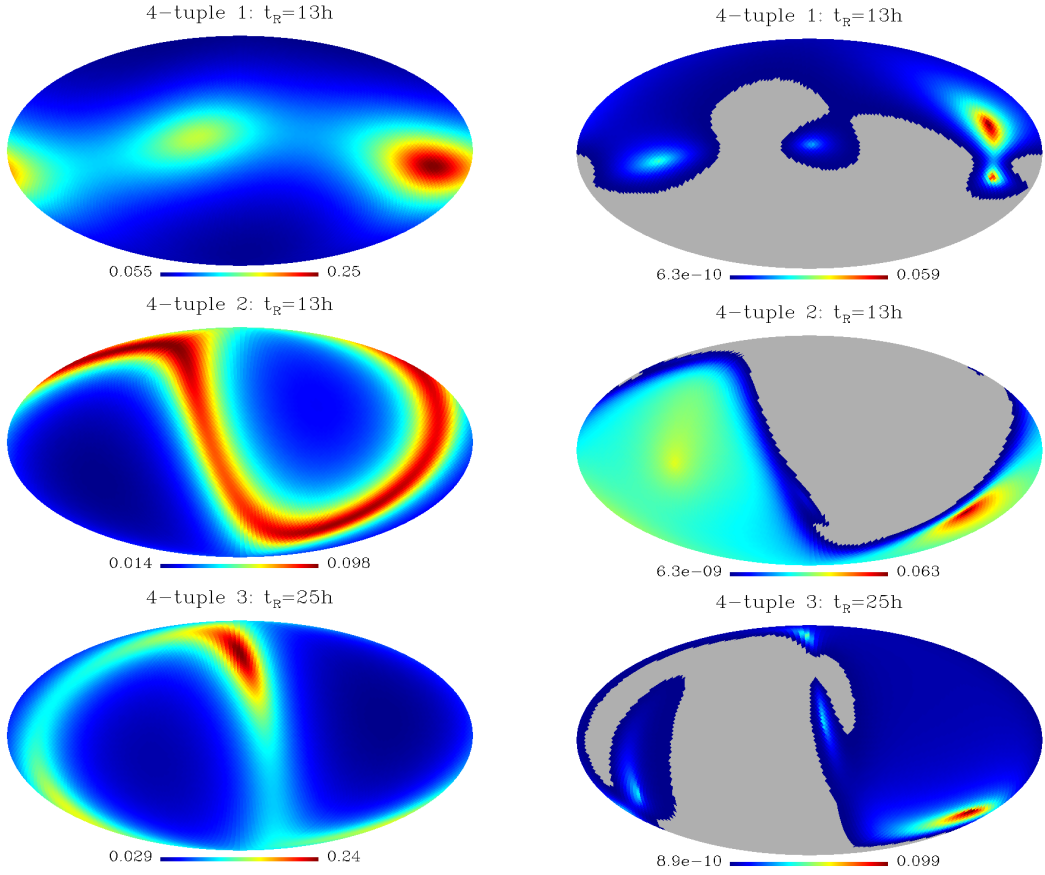}
}
\caption{All panels are HEALPIx mollwide colored representations.
Left (right) panels show the time distance 
$\lambda_{-}$ ($\lambda_{max} - \lambda_{-}$) for each direction on the hypersurface 
$\tau^{4} = \tau_{e}^{4}$. The $\lambda_{-}$ values define a 2-surface where 
$\chi^{2}$ vanishes. Each point inside this surface leads to a single valued 
positioning in physical space-time. The gray zones of the right panels 
correspond to the directions for which there are no positioning 
solutions for $\lambda > \lambda_{-}$. The colored part of these 
panels gives the width of the zone -surrounding the surface $\chi^{2}=0$--
whose points lead to double valued positioning.
Time $t_{R} $ and the 4-tuple of satellites used for positioning are
given above each panel.}
\label{figu3} 
\end{center}      
\end{figure*}

\begin{figure*}[tb]
\begin{center}
\resizebox{0.9\textwidth}{!}{%
\includegraphics{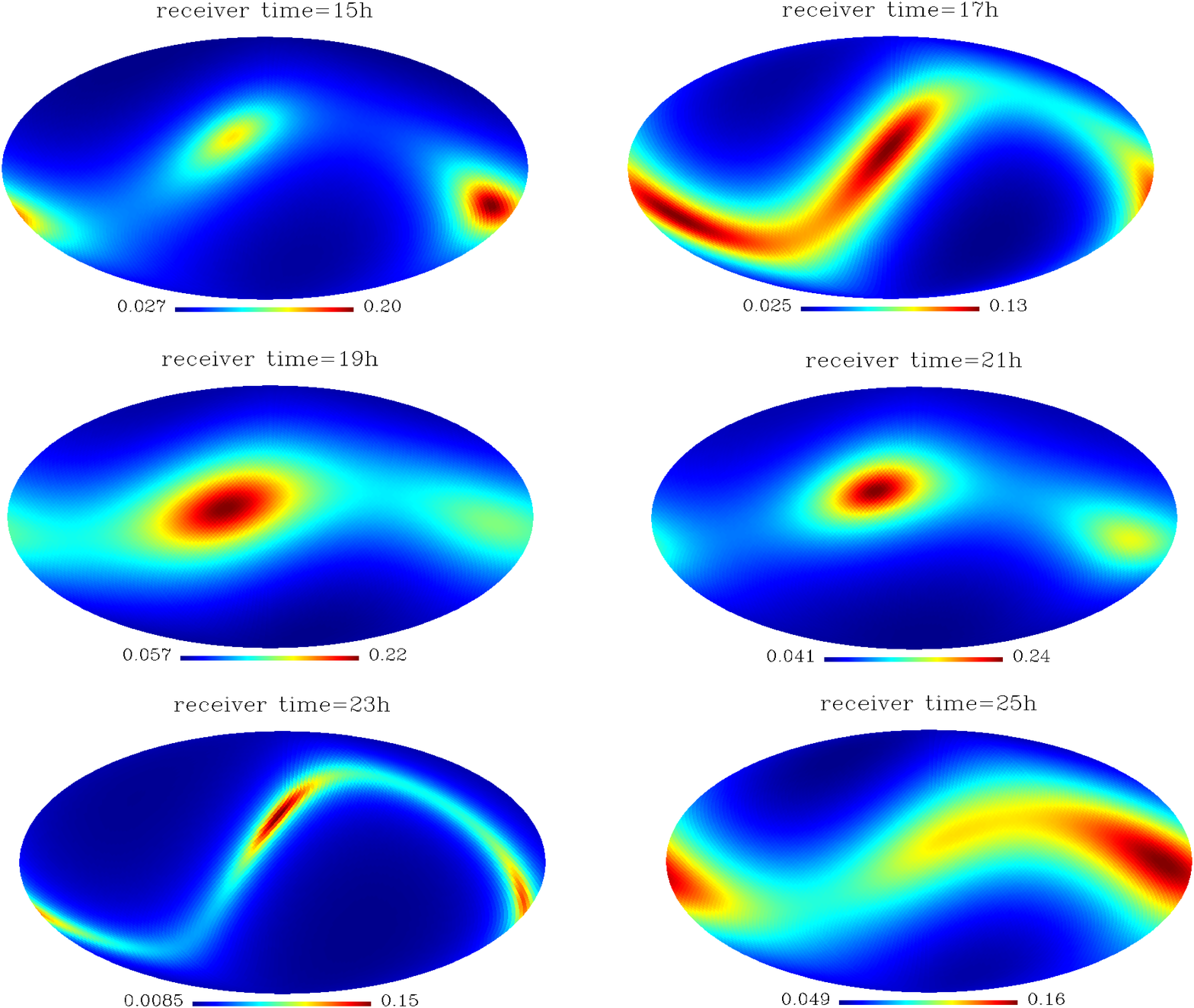}
}
\caption{Same as in the left panels of Fig.~\ref{figu3}
for the 4-tuple 1 and the same $t_{R} $ times as in Fig.~\ref{figu2}}
\label{figu4} 
\end{center}      
\end{figure*}

\begin{figure*}[tb]
\begin{center}
\resizebox{0.9\textwidth}{!}{%
\includegraphics{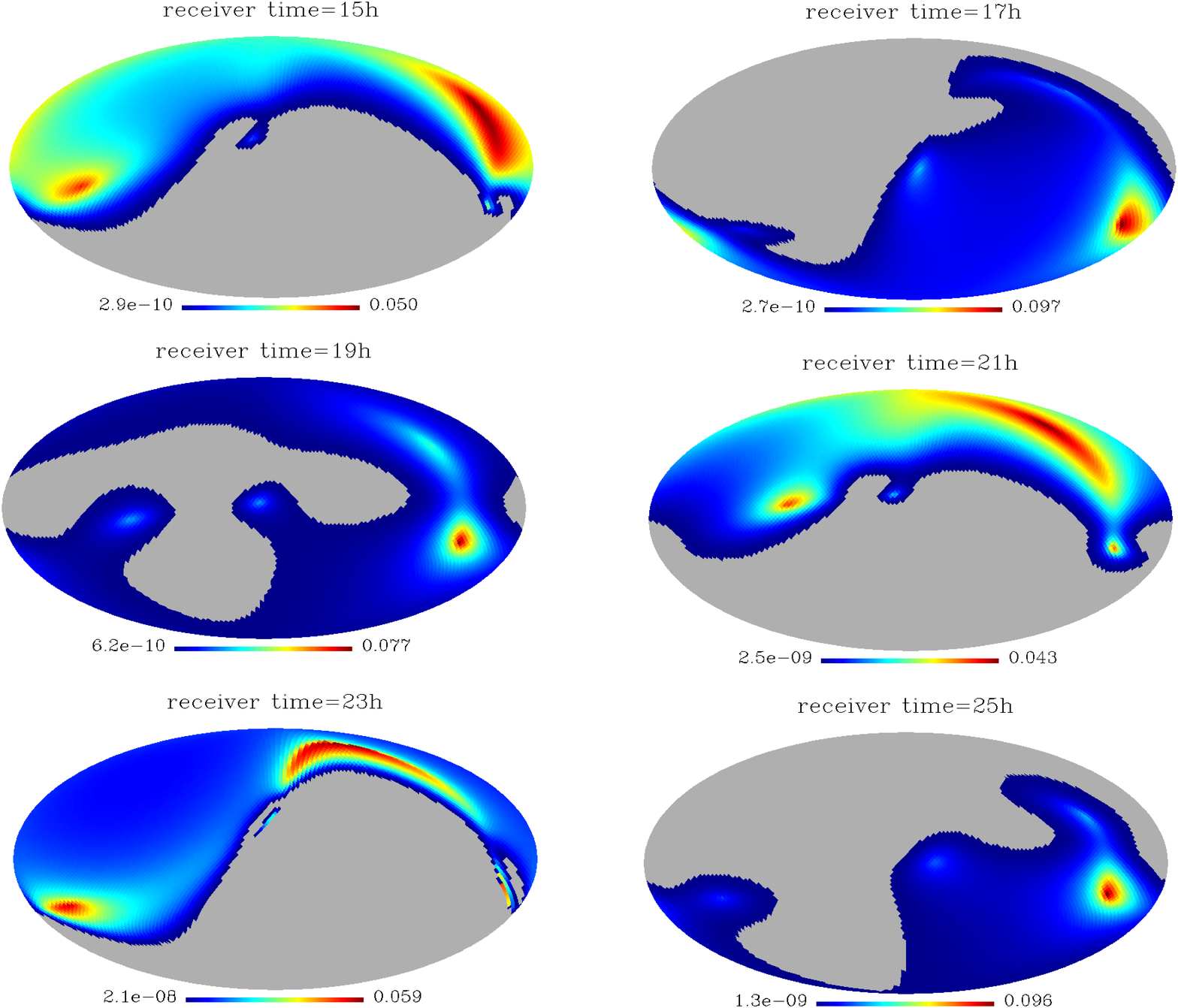}
}
\caption{Same as in the right panels of Fig.~\ref{figu3}
for the 4-tuple 1 and the same $t_{R} $ times as in Fig.~\ref{figu2}}
\label{figu5} 
\end{center}      
\end{figure*}

According to \cite{col10}, the sign of $\chi^{2} $ is 
crucial to characterize single and double positioning. Since $\chi $ is a vector
orthogonal to the hyperplane containing the four emission events, 
this hyperplane is time-like, null, and space-like for
$\chi^{2} >0$, $\chi^{2} =0$, and $\chi^{2} <0$, respectively. 
Single positioning corresponds to $\chi^{2} \leq 0$ (space-like and null hyperplanes), 
whereas location bifurcations appears for $\chi^{2} >0$ (time-like hyperplanes). 
Once the emission coordinates are known, the internal satellite configuration and the 
sign of $\chi^{2} $ may be easily obtained. Thus, the location bifurcations (double 
positioning, $\chi^{2} >0$) may be found. In the double positioning cases,
besides the emission coordinates, other measurements --either angles or times--
have been proposed (see Sect.~\ref{sec-3}) to 
identify the true location between the two possible ones. On account of these facts, 
we have found the zones in the emission coordinate region and co-region corresponding to
single and double positioning. Results have been presented in Figs.~\ref{figu1}
to~\ref{figu5}. In any case, there is a central zone with $\chi^{2} \leq 0$,
and a complementary one corresponding to double positioning.

\begin{figure*}[tb] 
\begin{center}
\resizebox{0.6\textwidth}{!}{%
\includegraphics{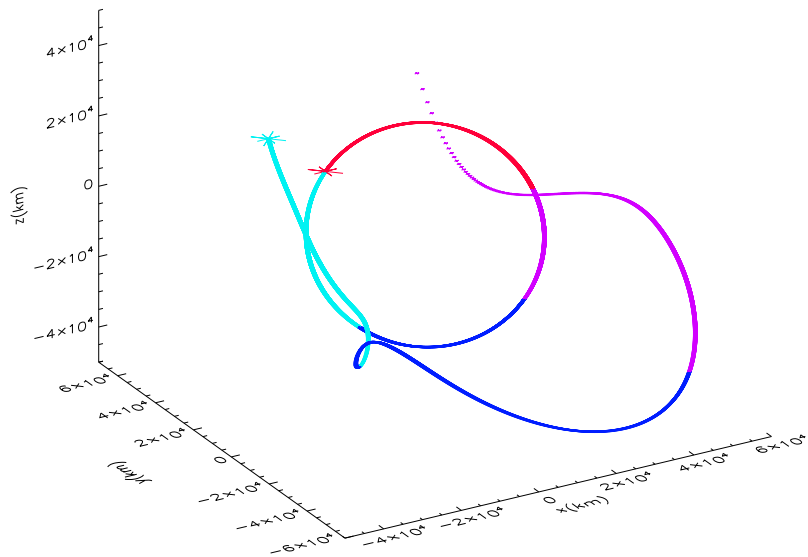}
}
\caption{Positioning a Galileo satellite with four GPS
emitters. The red star is a single valued initial point. It is 
the first point of a red arc of single valued solutions. Going through 
the circumference in the dextrogyre sense, we find a second
continuous arc of double valued positions successively colored in fuchsia, dark blue and 
and light blue. This arc returns to the initial point. The line of the 
associated positions is overrun in the same sense. It tends to infinity 
in the transition from the fuchsia to the red arcs. The final positioning
is double valued. One of the positions coincides with the initial one and 
the associated position is represented by the light blue star. }
\label{figu6} 
\end{center}      
\end{figure*}

\begin{figure*}[tb] 
\begin{center}
\resizebox{0.6\textwidth}{!}{%
\includegraphics{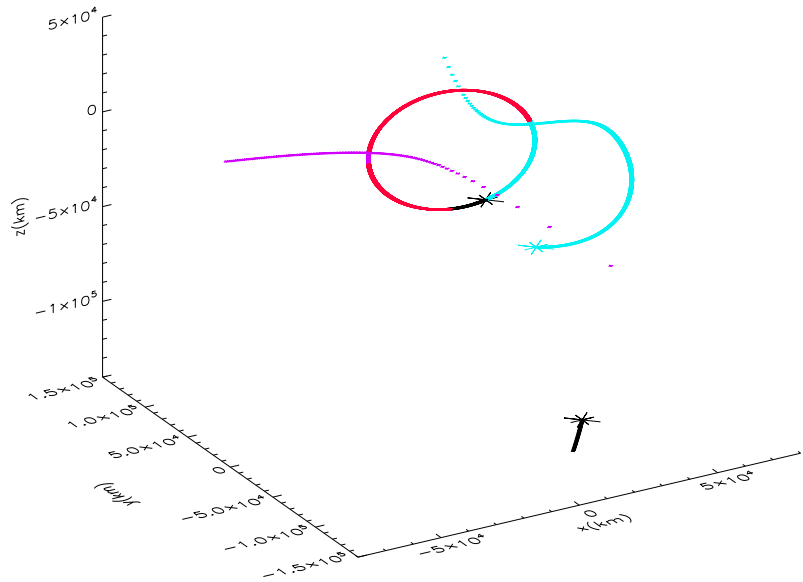}
}
\caption{Positioning a Galileo satellite with four GPS
emitters. The two black stars are the initial points  
of a double valued solution. The following succession of arcs is observed 
on the circumference: black, red, fuchsia, red, light blue. 
The corresponding lines outside the circumference tend to infinity in the 
following transitions: 
from black to red, from fuchsia to red, and from light blue to red.  
As in Fig.~\ref{figu6}, The light blue star 
is the final position. It is associated to the initial position on the
circumference.}
\label{figu7} 
\end{center}      
\end{figure*}

\begin{figure*}[tb]
\begin{center}
\resizebox{0.6\textwidth}{!}{%
\includegraphics{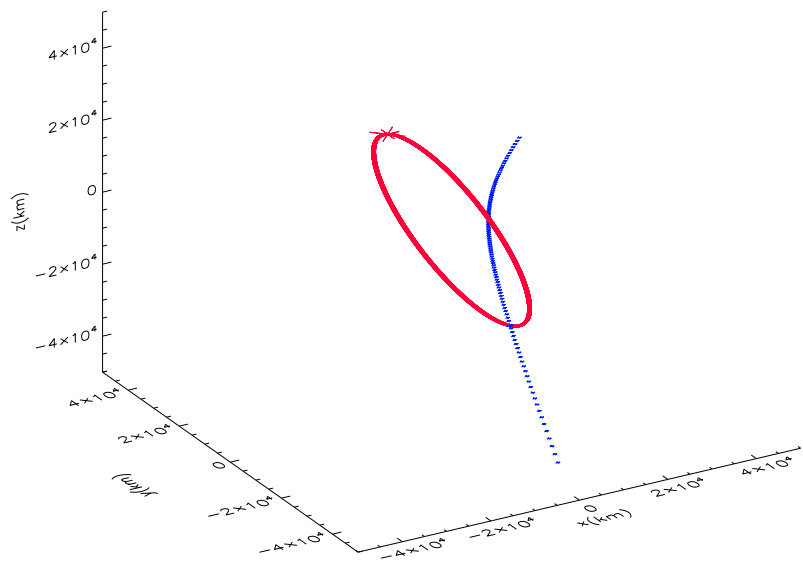}
}
\caption{Positioning a GPS satellite with four Galileo
emitters. The red star is the single valued initial position and 
also the final one. All the positions are single valued excepting a very small
dark blue arc on the circumference. The corresponding curve outside the circumference 
tends to infinity in the two transitions from dark blue to red.}
\label{figu8} 
\end{center}      
\end{figure*}

\begin{figure*}[tb]
\begin{center}
\resizebox{0.6\textwidth}{!}{%
\includegraphics{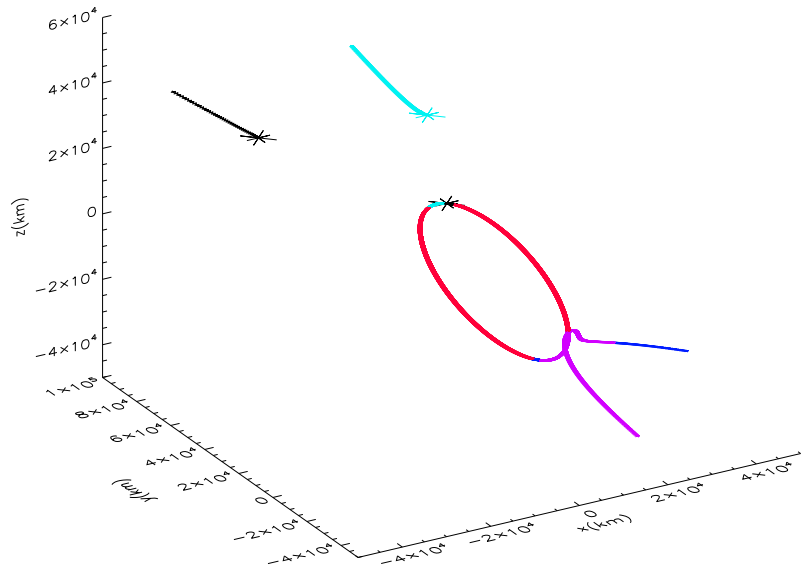}
}
\caption{Positioning a GPS satellite with four Galileo
emitters. The two black stars correspond to a double valued initial 
position. On the circumference we find the following succession of arcs: black, red,
fuchsia, dark blue, red, and light blue. The curves outside the circumference
tend to infinity in the following transitions: from black to red, from fuchsia 
to red, from dark blue to red, and from light blue to red. The final point 
is represented by the light blue star associated to the initial position 
located on the circumference}
\label{figu9} 
\end{center}      
\end{figure*}

If a receiver (for example a satellite) is always located inside the central region
of four emitter satellites, its positioning is single, and the emission coordinates 
are sufficient to find the receiver position; however, if the receiver enters
and leaves the central zone either one or various times, there are phases of 
single and double positioning. These are the cases represented in 
Figs.~\ref{figu6} to~\ref{figu9},
in which, a satellite of a certain GNSS is positioned by using four satellites of
other GNSS. A typical orbit of the GPS and Galileo satellite constellations
enters and leaves various times the central region. This means that there are 
bifurcation locations and, consequently, apart from the emission coordinates, 
devices measuring either angles or times are needed to get the true 
receiver position.

A clock aboard of the receiver satellite could measure the observational 
time, $t_{o}$, when the emission coordinates are received. This 
measurement may be done 
whatever the positioning character may be. In the case of single positioning, 
there is only a coordinate time $t_{1} $ derived from the 
emission coordinates, whereas a pair of coordinate times ($t_{1},t_{2}$) appears 
in any double positioning.
In the absence of errors in both the coordinate times and the observational 
time $t_{o}$, this last time would exactly coincide with $t_{1} $ for 
single positioning, and with one of the times ($t_{1},t_{2}$) in the 
case of double positioning. The time coinciding with $t_{o} $ would 
correspond to the true location.
These coincidences would not be 
exact due to both the limited accuracy of the clock measuring $t_{o} $, and
the positioning errors in $t_{1}$ or in the pair ($t_{1},t_{2}$). 
These errors may be associated, for example, to uncertainties 
in the world lines of the four
emitters. The cases of single positioning might be used to calibrate the errors
separating $t_{1} $ from $t_{0} $. For double positioning 
the differences $|t_{1} - t_{0}| $ and $|t_{2} - t_{0}| $ must be first
estimated and, then, two cases may be distinguished: (i) 
quantity $|t_{1}-t_{2}|$ is much larger than the typical 
value taken by $|t_{1} - t_{0}| $
in single positioning, and (ii) the absolute value 
$|t_{1}-t_{2}|$ is of the order of the mentioned typical value.
In the first case, the smallest of the $|t_{1} - t_{0}| $ and $|t_{2} - t_{0}| $
quantities clearly corresponds to the true positioning; however, in case (ii),
the true position may not be found with the clock on board. In such a case,
it could be studied if the criterion based on angle measurements (see Sect.~\ref{sec-3})  
may lead to better results.

We have estimated the maximum and minimum values of $|t_{1}-t_{2}|$ 
for the double positioning events appearing in Figs.~\ref{figu6} to~\ref{figu9}.
The maximum value is $|t_{1}-t_{2}|_{max} \simeq 128.6 \ s$. In general, large 
values of $|t_{1}-t_{2}|$ appear close to the transitions 
from double to single positioning zones. 
There are no problems to choose the right position is these cases.
The minimum value is $|t_{1}-t_{2}|_{min} \simeq 1.1  \times 10^{-6} \ s $;
it is also large as compared to the expected errors in the time $t_{o} $ measured by 
a good atomic clock (aboard the receptor satellite). The errors 
due to uncertainties in the satellite orbits strongly depends 
on the Jacobian of the transformation from 
emission to inertial coordinates \citep{neu10}. This Jacobian 
is to be calculated for the values of the emission coordinates
corresponding to the double positioning under consideration.
The detailed study of this type of errors as well as the analysis of other possible 
error sources are open problems requiring further research.

\acknowledgments{We would like to thank J. A. Morales for
valuable discussions. This work has been supported by the Spanish
Ministerio de Ciencia e Innovaci\'on, MICINN-FEDER project
FIS2009-07705.}

\end{document}